\documentclass[aps,prl,twocolumn,superscriptaddress,amsmath,amssymb]{revtex4-1}

\usepackage{graphicx}
\usepackage{multirow}
\usepackage{natbib}
\usepackage{array}
\usepackage{multirow,amssymb,amsbsy,amsmath,epstopdf}
\usepackage{color}

\newcommand{\ket}[1]{\ensuremath{\left|#1\right\rangle}}
\usepackage[colorlinks,citecolor=blue]{hyperref}

\begin{document}
\title{Robust Device-Independent Characterization of Sharpness and Incompatibility of Unsharp Instruments}
\author{Qian Zhang}
\affiliation{College of Physics and Optoelectronic Engineering, Ocean University of China, Qingdao 266100, China.}
\author{Kai-Yu Yuan}
\affiliation{China Mobile (Suzhou) Software Technology Company Limited, Suzhou 215163, China.}
\author{Yan-Xin Rong}
\affiliation{College of Physics and Optoelectronic Engineering, Ocean University of China, Qingdao 266100, China.}
\author{Zhen Shang}
\affiliation{College of Physics and Optoelectronic Engineering, Ocean University of China, Qingdao 266100, China.}
\author{Yong-Jian Gu}\email{yjgu@ouc.edu.cn}
\affiliation{College of Physics and Optoelectronic Engineering, Ocean University of China, Qingdao 266100, China.}
\affiliation{Engineering Research Center of Advanced Marine Physical Instruments and Equipment (Ministry of Education), Ocean University of China, Qingdao 266100, China.}
\affiliation{Qingdao Key Laboratory of Advanced  Optoelectronics, Ocean University of China, Qingdao 266100, China}
\author{Ya Xiao}\email{xiaoya@ouc.edu.cn}
 \affiliation{College of Physics and Optoelectronic Engineering, Ocean University of China, Qingdao 266100, China.}
\affiliation{Engineering Research Center of Advanced Marine Physical Instruments and Equipment (Ministry of Education), Ocean University of China, Qingdao 266100, China.}
\affiliation{Qingdao Key Laboratory of Advanced  Optoelectronics, Ocean University of China, Qingdao 266100, China}
\begin{abstract}
Unsharp measurements are key resources for tasks that balance information gain and disturbance, but certifying them without device assumptions remains a challenge. We propose a fully device-independent protocol for characterizing unsharp instruments, based on an entanglement-assisted sequential quantum random access code, where the first decoder is allowed to communicate her measurement setting to the second. This communication-enhanced scheme creates a decoding regime in which both decoders surpass classical bounds, enabling tight quantification of sharpness and direct quantification of measurement incompatibility beyond noncommunicating protocols. Experimentally, we implement tunable unsharp measurements using a Mach-Zehnder interferometer, observing the predicted sequential enhancement in decoding probability. Additionally, we achieve significantly narrower sharpness intervals and incompatibility quantification across multiple target sharpness values. Our results show that communication is a powerful operational resource for certifying precisely unsharp instruments and advancing device-independent quantum information protocols.
 
\end{abstract}

\maketitle
	
\section{I. Introduction}
Unlike positive-operator-valued measures, which capture only the classical outcome of a measurement~\cite{davies1970operational,busch1986unsharp}, quantum instruments provide a complete operational description, including both the classical outcome and the resulting postmeasurement state~\cite{dressel2013quantum, khandelwal2025simulating}. This makes them essential for sequential measurement tasks, such as the characterization of quantum networks~\cite{chiribella2009theoretical}, studies of quantum causality~\cite{oreshkov2012quantum}, and the recovery of quantum correlations~\cite{kim2012protecting}. A particularly useful class of quantum instruments is unsharp instruments, also referred to as weak or unsharp measurements, which enable tunable measurement sharpness. This tunability establishes a trade-off between information gain and system disturbance~\cite{silva2015multiple}, underpinning a wide range of quantum information protocols, including sequential sharing of quantum correlations ranging from coherence~\cite{datta2018sharing} and contextuality~\cite{anwer2021noise,kumari2023sharing,xiao2025sharing}, to quantum entanglement~\cite{bera2018witnessing}, quantum steering~\cite{choi2020demonstration}, and Bell nonlocality~\cite{hu2018observation,maity2020detection}, as well as network nonlocality~\cite{mao2023recycling,sun2024network}, manipulation of steering direction \cite{han2022manipulating}, randomness certification ~\cite{foletto2020experimental,bowles2020bounding}, reduction in communication complexity~\cite{xiao2021widening,minati2026randomness}. A detailed review of these developments can be found in Ref.~\cite{cai2025review}.

Despite their importance, fully characterizing unsharp instruments remains a challenging task, as it requires the simultaneous certification of the underlying quantum states, processes, and measurements, together with the quantification of measurement sharpness and incompatibility. To address this challenge, Kliesch and Roth proposed protocols for quantum-state and quantum-process certification~\cite{kliesch2021theory}. Miklin \textit{et al.} introduced a tripartite sequential quantum random access code (QRAC) protocol, demonstrating that unsharp measurements cannot be simulated by projective measurements. This result established the theoretical feasibility of semi-device-independent (SDI) characterization of unsharp instruments~\cite{miklin2020semi}. Mohan \textit{et al.} further showed that sequential QRAC allow one to infer upper and lower bounds on a quantum instrument's sharpness parameter~\cite{mohan2019sequential}. Since sharpness directly determines an instrument's capability to sequentially share quantum correlations, certify extractable randomness, and achieve quantum advantages in communication tasks, establishing precise bounds is crucial. Recent advancements have extended these ideas to various prepare-transform-measure scenarios, enabling the SDI characterization of individual~\cite{ss2023robust} or multiple sharpness parameters~\cite{paul2024self,mukherjee2021semi}, with experimental validations in photonic systems~\cite{anwer2020experimental,foletto2020experimental}.

While SDI approaches have advanced the characterization of unsharp instruments, they inherently assume a bounded system dimension and cannot achieve full device independence. Recent device-independent (DI) protocols have begun to close this gap. Wagner \textit{et al.} proposed a DI characterization of quantum instruments~\cite{wagner2020device}, and Zhou \textit{et al.} experimentally certified quantum instrument using an improved DI scheme with enhanced robustness to noise~\cite{zhou2025experimental}. Roy and Pan introduced DI self-testing schemes to further quantify sharpness parameters via sequential Bell tests \cite{roy2023device}, while our previous work experimentally quantified sharpness bounds using entanglement-assisted sequential QRAC~\cite{xiao2021widening}. However, existing SDI and DI schemes exhibit reduced precision in quantifying sharpness as it decreases, and DI schemes  do not address measurement incompatibility-a cornerstone of quantum information theory and a crucial resource for quantum cryptography, state discrimination, and the verification of quantum correlations~\cite{buscemi2020complete,guhne2023colloquium}.

Here, we extend the protocol of Roy \textit{et al.}~\cite{roy2026robust} and present a DI protocol for characterizing unsharp instruments based on an entanglement-assisted sequential $2\rightarrow 1$ QRAC with one encoder and two decoders.  A key innovation is allowing the first decoder to communicate her measurement setting to the second, increasing the average decoding success probability beyond classical
limits. By deriving the trade-off between the decoders’ success probabilities, our protocol enables the simultaneous quantification of both measurement sharpness and incompatibility. Experimentally, we implement the protocol using a photonic setup, achieving significantly narrower intervals for the sharpness parameter compared to previous noncommunicating SDI and DI schemes. Additionally, we robustly quantify measurement incompatibility across multiple target sharpness values. Our work marks a significant step forward in the characterization of unsharp instruments and highlights communication as a different resource for the development of more precise and robust quantum information protocols.

\section{II. Scenario and Theory}
 
\begin{figure}[htbp]
	\centering
	\includegraphics[width=0.9\linewidth]{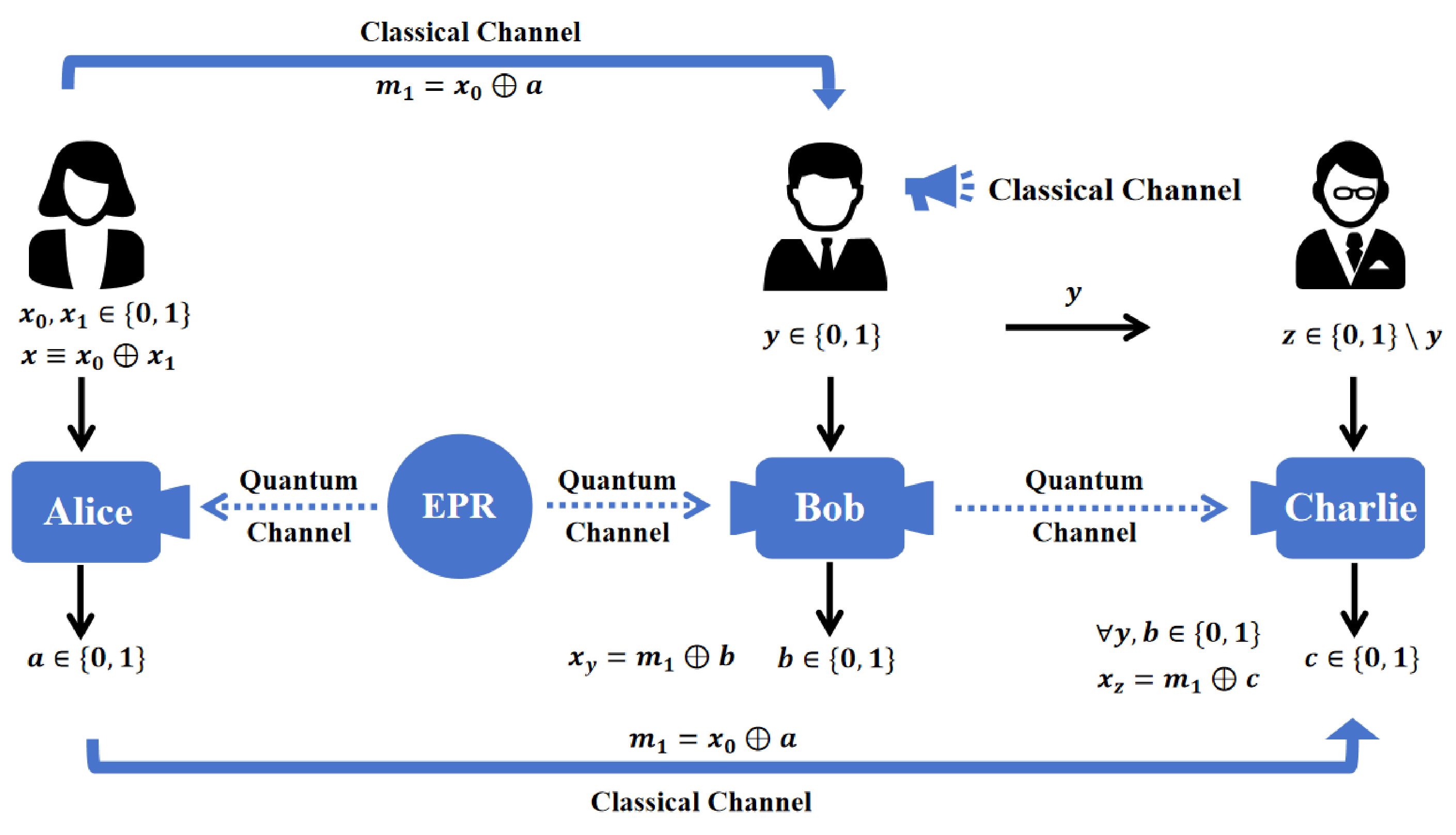}
	\caption{DI characterization of unsharp instruments via a sequential communication game. The schematic shows the entanglement-assisted QRAC protocol, which features sequential measurements and limited communication between the two decoders, Bob and Charlie. Bob is allowed to communicate his measurement setting to Charlie, enhancing the average decoding success probability beyond classical limits and enabling the simultaneous certification of both measurement sharpness and incompatibility.}
	\label{fig:1}
\end{figure}

The noise-resistant and device-independent characterization of unsharp instruments, implemented within an entanglement-assisted sequential $2\rightarrow 1$ QRAC with restricted communication, is illustrated in Fig.~\ref{fig:1}. The protocol involves one encoding party, Alice, and two decoding parties, Bob and Charlie. Initially, Alice and Bob share a two-qubit entangled state $ \rho_{AB} $. Alice then encodes her classical inputs $ x_{0}, x_{1}\in \{0,1\}$ by performing a local measurement $A_{a\vert\vec{r}_{x}} =(I+(-1)^a \vec{r}_{x} \cdot \vec{\sigma})/2$ along the $\vec{r}_{x}$ direction on one side of the shared state. The measurement setting is selected according to $ x= x_{0}\oplus x_{1}$. The resulting outcome $a \in \{0,1\}$ allows Alice to generate the classical message $ m_{1}= x_{0}\oplus a $, which she forwards independently to Bob and Charlie to improve their decoding success probabilities. Bob then performs an unsharp measurement $ B_{b\vert\vec{s}_{y}}=K^{\dagger}_{b\vert\vec{s}_{y}}K_{b\vert\vec{s}_{y}}=(I+(-1)^b \eta\vec{s}_{y} \cdot \vec{\sigma})/2$ along the $\vec{s}_{y}$ direction on the other side of the shared state, based on an input $y \in \{0,1\}$,  to decode the desired bit $ x_{y}$. Here, $\eta$ and $ K_{b\vert\vec{s}_{y}}$ represent the measurement sharpness and the corresponding Kraus operator, respectively. Bob records the output $b \in \{0,1\}$ and compares it with the received bit $ m_{1} $. The decoding process is successful if $ x_{y}= m_{1}\oplus b $, and the success probability can be calculated by $P(x_{y}= m_{1}\oplus b\vert x,y)=\text{Tr}[ A_{a\vert\vec{r}_{x}}\otimes B_{b\vert\vec{s}_{y}}\rho_{AB}]$. In fact, Bob's operations are described by the notion of a quantum instrument~\cite{davies1970operational}, which produces a classical measurement outcome and a corresponding post-measurement state.
Here, Bob not only relays the postmeasurement state as in previous protocols~\cite{tavakoli2015quantum}, but also transmits his measurement setting $y$ to Charlie. Similarly, Charlie receives an input $z \in \{0,1\}$ with the aim of guessing $ x_{z}$  by performing an optimal measurement $ C_{c\vert\vec{t}_{z}} =(I+(-1)^c \vec{t}_{z} \cdot \vec{\sigma})/2$ along the$\vec{t}_{z}$ direction on the state he receives from Bob, yielding an outcome $c \in \{0,1\}$. As the communication is restricted, Charlie only has the information about Bob's measurement setting, but not about the corresponding outputs. Thus, for a given $y$, we average over the measurement outputs to obtain the state shared between Alice and Charlie, which is given by $\rho_{AC}= \dfrac{1}{2}\sum \limits_{b}( I_{A}\otimes K_{b\vert\vec{s}_{y}})\rho_{AB}( I_{A}\otimes K^{\dagger}_{b\vert\vec{s}_{y}})$. Charlie's success probability is then given by $P(x_{z}= m_{1}\oplus c\vert x,z)=\text{Tr}[ M_{a\vert\vec{r}_{x}}\otimes M_{c\vert\vec{t}_{z}}\cdot\rho_{AC}]$. Assuming all inputs $(x_{0},x_{1},y,z)$ are statistically independent and uniformly distributed, the average success probabilities of Bob and Charlie, $P_{AB} $ and $ P_{AC}$, are given by 
\begin{equation}\label{pro1}
\begin{split}
&P_{AB}=\dfrac{1}{8}\sum \limits_{x,y}P(x_{y}= m_{1}\oplus b\vert x,y), \\
&P_{AC}=\dfrac{1}{8}\sum \limits_{x,z}P(x_{z}= m_{1}\oplus c\vert x,z).\\
\end{split}
\end{equation}

When the shared state is the Bell state $|\psi\rangle = (|01\rangle + |10\rangle)/\sqrt{2}$, the optimal measurement directions are $\vec{r}_{0}=\{1,0,0\}$ and $ \vec{r}_{1}=\{0,0,-1\}$ for Alice, $\vec{s}_{0}=\vec{y}_{0}=\{1/\sqrt{2},1/\sqrt{2},0\}$ and $ \vec{s}_{1}=\vec{t}_{1}=\{1/\sqrt{2},-1/\sqrt{2},0\}$ for Bob and Charlie. This yields the maximum average success probabilities $P_{AB} = \left(2 + \sqrt{2}\eta\right)/4$ and $P_{AC} = \left(2 + \sqrt{2-2\eta^2}\right)/4$. In our communication scenario, Bob's role is similar to that of the traditional RAC. The optimal classical limit of Bob's average success probability is $3/4$ \cite{tavakoli2015quantum}. Since Charlie has the information about Bob's measurement setting, his success probability remains limited to $1/2$ \cite{roy2026robust}. By adjusting  $\eta \in (1/\sqrt{2}, 1)$, both $P_{AB}$ and $P_{AC}$ can exceed their classical bounds. Furthermore, by eliminating $\eta$ from $P_{AB}$ and $P_{AC}$, one can obtain the following trade-off relation
\begin{equation} \label{trade-off}  
	P_{AC} = \frac{1}{2} + \sqrt{P_{AB}-P_{AB}^2-\frac{1}{8}}.
\end{equation}
Although Eq.~\eqref{trade-off} provides the optimal value of $P_{AC}$ for a given value of $P_{AB}$ that uniquely implies a precise value of $\eta$ in the ideal scenario, experimental data are typically suboptimal. Nevertheless, the experimentally obtained values of $P_{AB}$ and $P_{AC}$ can still yield rigorous upper and lower bounds on $\eta$  by

\begin{equation}
	\begin{aligned}
	\eta \geq \sqrt{2}(2P_{AB}-1) \equiv \eta_{\text{min}},\\
	 \quad \eta \leq \sqrt{8P_{AC}-8(P_{AC})^2-1} \equiv \eta_{\text{max}}.
		\end{aligned}
\end{equation}
Clearly, as the experimentally observed success probabilities approach the optimum given by Eq.~(\ref{trade-off}), the interval between $\eta_{\text{min}}$ and $\eta_{\text{max}}$ narrows, yielding a more precise estimation of $\eta$.

Since certification of quantum advantage requires the use of incompatible measurements, the success probabilities $P_{AB}$ and $P_{AC}$ can also be employed to quantify the incompatibility of measurements. The incompatibility of two measurements along $\vec{n}_0$ and $\vec{n}_1$ can be quantified as $D(\vec{n}_0, \vec{n}_1) = |\vec{n}_0 + \vec{n}_1| + |\vec{n}_0 - \vec{n}_1| - 2$ ~\cite{busch2014heisenberg}. 
Incompatible measurements satisfy $D(\vec{n}_0, \vec{n}_1)>0$. In the Appendix, we further show that the incompatibility of Bob's pair of unsharp measurements and that of Charlie's pair of projective measurements are respectively bounded by
\begin{equation}
\label{eq:10}
\begin{split}
 &D_B \geq \max  \{ 8P_{AB} - 6, 0  \}, \\
 &D_C \geq \max \{ \frac{8P_{AC} - 4}{\sqrt{4P_{AC}^2 - 4P_{AC} + 3/2}} - 2, 0 \}.
\end{split}
\end{equation}
 
\section{III. Experimental Setup}
 \begin{figure}[htbp]
 \centering\includegraphics[width=8.5cm]{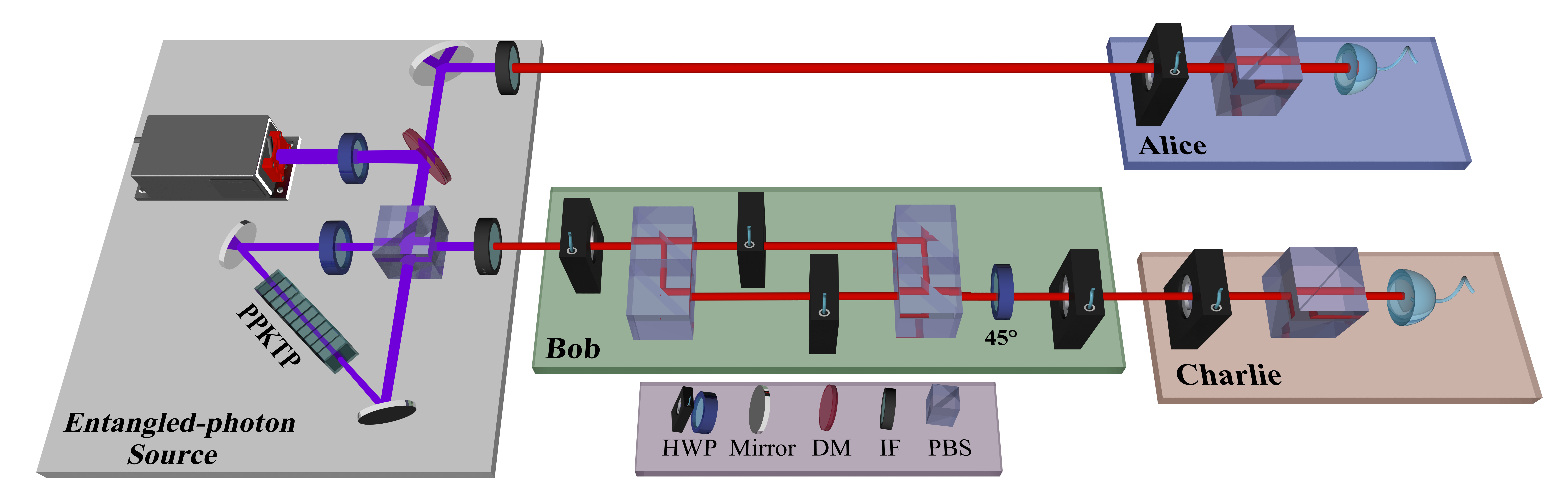} 
	\caption{Experimental setup. The apparatus comprises four functional modules: the gray module for polarization-entangled photon-pair generation; the purple module for Alice's projective measurement; the green module for Bob's unsharp measurement; and the orange module for Charlie's projective measurement. Abbreviations: HWP, half-wave plate; DM, dichroic mirror; PBS, polarizing beam splitter; PPKTP, periodically poled $\text{KTiOPO}_{4} $ crystal; IF, interference filter. 
}
	\label{fig:setup}  
\end{figure} 

Our experimental setup is depicted in Fig.~\ref{fig:setup}. Polarization-entangled photon pairs are generated via type-II spontaneous parametric down-conversion in a periodically poled $\text{KTiOPO}_{4}$ (PPKTP) crystal placed in a Sagnac interferometer. A diagonally polarized pump beam is split into horizontally (transmitted) and vertically (reflected) polarized components by a dual-wavelength polarizing beam splitter (PBS), which then pump the PPKTP crystal along opposite directions in the Sagnac ring. A dual-wavelength half-wave plate (HWP) set at $45^{\circ}$ in the reflected port rotates the polarization of the reflected pump beam to horizontal for efficient down-conversion, while also rotating the polarization of the down-converted photons in the transmitted path by $90^{\circ}$ to compensate temporal walk-off. The two counter-propagating down-converted photons are recombined at the dual-wavelength PBS, erasing any distinguishing path information. After filtering out the residual pump light with two 3-nm-bandwidth interference filters (IFs), the resulting two-qubit polarization-entangled state $(|HV\rangle + |VH\rangle)/\sqrt{2}$ is obtained. Here, the horizontal and vertical polarization states $\ket{H}$ and $\ket{V}$ encode the logical states $\ket{0}$ and $\ket{1}$, respectively. Quantum state tomography yields a state fidelity of $0.98\pm 0.02$ and a concurrence of $0.97\pm 0.01$, confirming the high quality of the generated state.

One photon from each entangled pair is sent to Alice for encoding, while the other is sent sequentially to Bob and Charlie for decoding. Alice encodes classical bits via a projective measurement implemented with a HWP and a PBS (purple module). Specifically, the HWP is set at $22.5^\circ$ for $A_{0|0}$, $-22.5^\circ$ for $A_{1|0}$, $45^\circ$ for $A_{0|1}$, and $0^\circ$ for $A_{1|1}$.  Bob and Charlie then perform the corresponding optimal measurements to decode the desired bit. To simulate the communication between Bob and Charlie, we use a random number generator to produce the required input, which controls their HWPs to perform the desired measurements.

Bob's unsharp instrument is implemented using a tunable Mach-Zehnder interferometer, as shown in the green module of Fig. \ref{fig:setup}. Two glued PBSs transmit the $\ket{H}$ polarization and laterally shift the $\ket{V}$ component by 3 mm. The HWP in the $\ket{H}$ path is set at $45^\circ - \theta$, the HWP in the $\ket{V}$ path at $\theta$, and an additional HWP after the second PBS is set at $45^\circ$, realizing the unsharp measurement $\eta \sigma_z$, with sharpness $\eta = \cos(4\theta)$ and output $b=0$.  To implement an arbitrary unsharp measurement $M_{b|\vec{s}_y}$, a basis transformation is achieved using two additional HWPs set to the same angle. The HWP before the first glued PBS maps the eigenstates of $\vec{s}_y \cdot \vec{\sigma}$ onto those of $\sigma_z$, while the HWP after the second glued PBS performs the inverse transformation (see Refs.~\cite{hu2018observation,xiao2021widening} for details). The desired measurements $\{B_{0|0}, B_{1|0}, B_{0|1}, B_{1|1}\}$ are realized by setting HWP2 and HWP4 to the corresponding angles $\{11.25^\circ, -33.75^\circ, 33.75^\circ, -11.25^\circ\}$. 
 
Charlie's decoding operations are similar to Bob's but replace the unsharp measurement with a projective one. The measurements $C_{0|0}$, $C_{1|0}$, $C_{0|1}$, and $C_{1|1}$ are implemented using a HWP followed by a PBS (orange module), with the HWP set at $11.25^\circ$, $-33.75^\circ$, $33.75^\circ$, and $-11.25^\circ$, respectively.

In the experiment, nine distinct values of $\eta$ were assigned: $\eta\!=\!\{$0.31, 0.44, 0.56,  0.67, 0.77, 0.85, 0.91, 0.96, 0.99$\}$. Photons were delivered via single-mode fibers to the single-photon avalanche diodes at Alice's and Charlie's sides, providing coincidence counts used to estimate the success probabilities $P_{AB}$ and $P_{AC}$ for each $\eta$. These probabilities were subsequently employed to characterize the sharpness of Bob's measurements and to evaluate the degree of incompatibility of the pair of measurements performed by Bob and Charlie, respectively.

\section{IV. Experimental Results}
\begin{figure}[htbp]
	\centering\includegraphics[width=7.5cm]{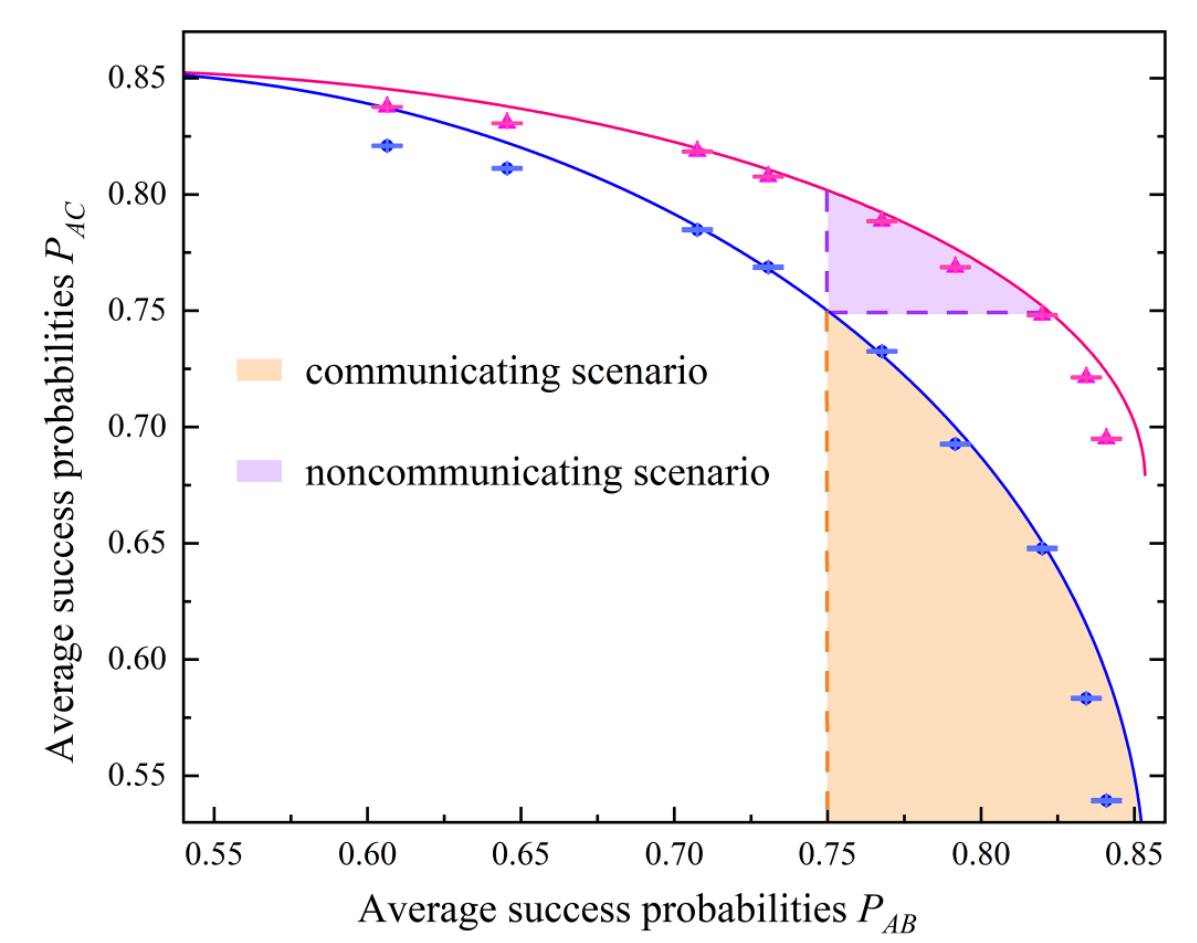} 
	\caption{Trade-off between average success probabilities $P_{AB}$ and $P_{AC}$ as a function of sharpness parameter $\eta$. Pink and Blue curves show the theoretical optimal trade-offs in the entanglement-assisted sequential QRAC scenario without and with classical communication between the decoders, respectively. Corresponding experimental values are marked as pink triangles and blue points.  Error bars are estimated from Poisson counting statistics.  Shaded purple and orange regions indicate where both $P_{AB}$ and $P_{AC}$ exceed their respective classical bounds in the noncommunicating and communicating scenarios, characterized by the conditions ($P_{AB}>3/4$, $P_{AC}>3/4$) and ($P_{AB}>3/4$,  $P_{AC}>1/2$), respectively.}
	\label{fig:3}  
\end{figure}

\begin{figure}[htbp]
	\centering\includegraphics[width=7.5cm]{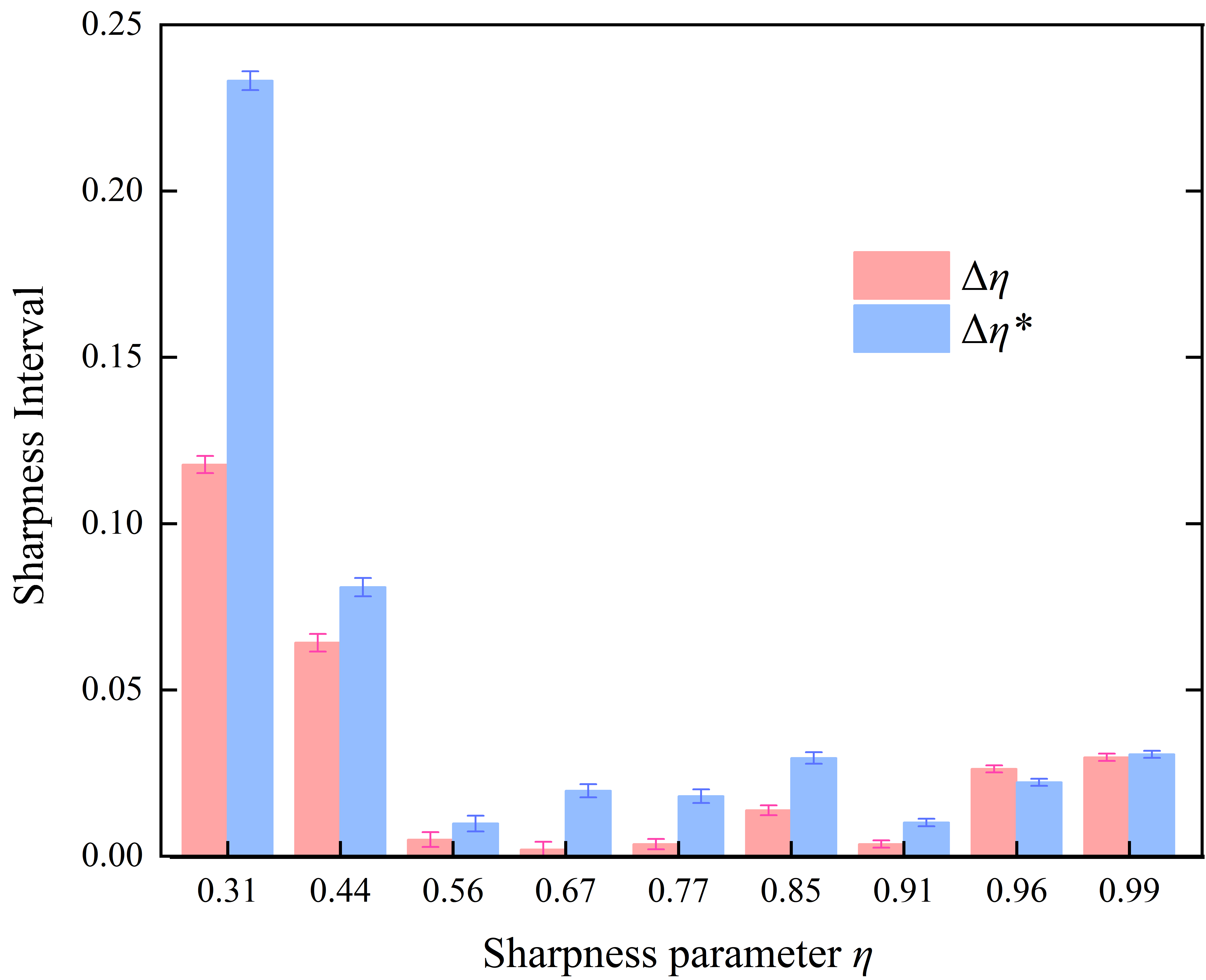} 
	\caption{The sharpness intervals, defined as the differences between the upper and lower bounds of Bob's sharpness parameter $\eta$, are shown for nine target values. $\Delta \eta^*$ (blue) and $\Delta \eta$ (pink) represent the sharpness intervals obtained in the noncommunicating and communicating scenarios, respectively. Error bars are estimated from Poissonian counting statistics.}
	\label{fig:4}  
\end{figure}  

\begin{figure}[htbp]
	\centering\includegraphics[width=7.5cm]{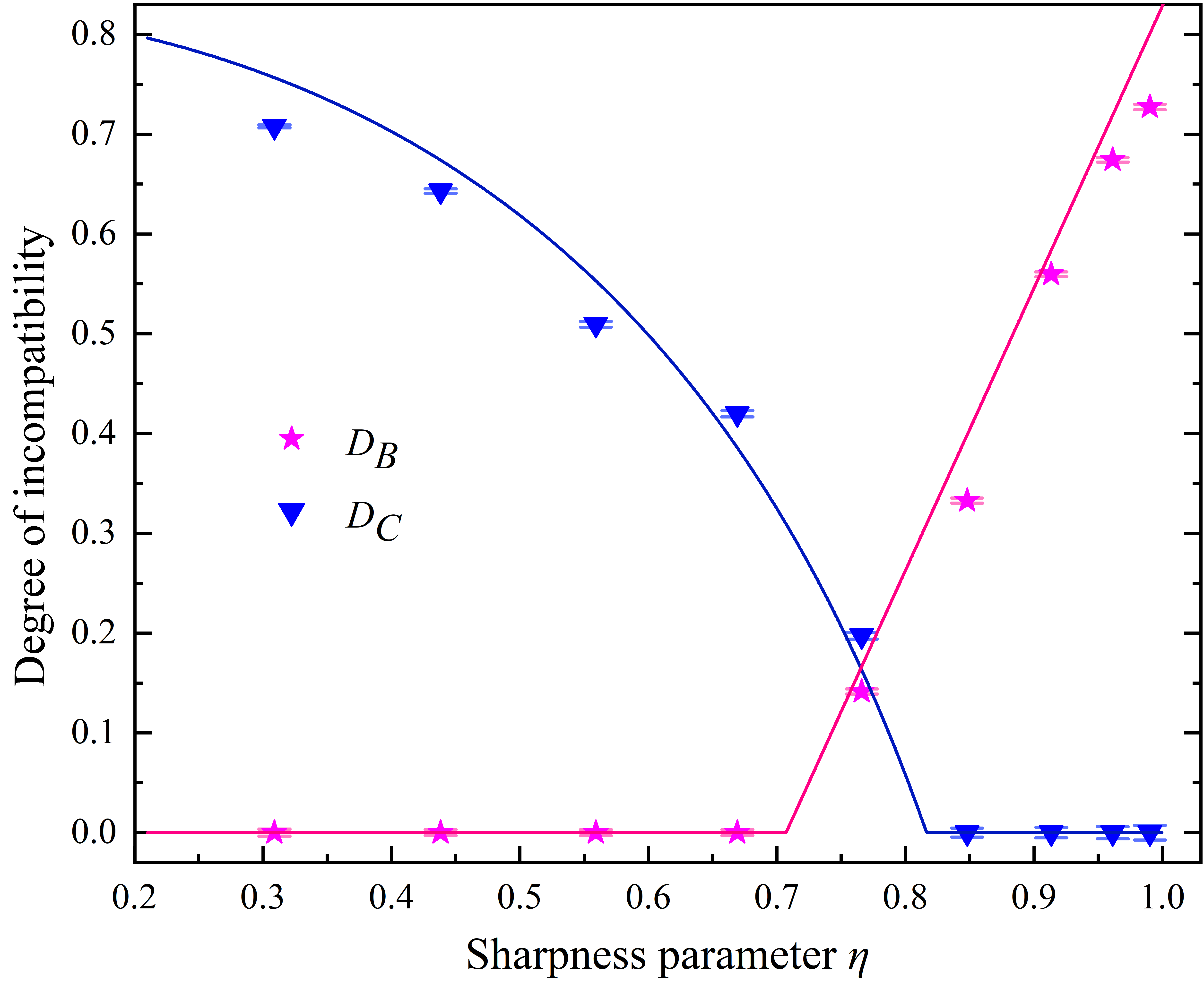} 
	\caption{Lower bound on the degree of incompatibility in Bob's (pink) and Charlie's (blue) respective pair of measurements for nine targeted values of the sharpness parameter $\eta$. Error bars are derived from Poissonian counting statistics.}
	\label{fig:5}  
\end{figure} 

To demonstrate the advantage of our communication protocol, we compare it with the standard entanglement-assisted sequential QRAC scenario in which decoders do not communicate (experimental data for this case are taken from Ref.~\cite{xiao2021widening}). The trade-off between the average success probabilities $P_{AB}$ and $P_{AC}$ is shown in Fig.~\ref{fig:3}. The optimal trade-offs are represented by the blue and pink curves for the communicating and noncommunicating entanglement-assisted sequential QRAC, respectively. Shaded purple and orange regions indicate where both $P_{AB}$ and $P_{AC}$ outperform their corresponding classical bounds— namely, $P_{AB} > 3/4$ and $P_{AC} > 3/4$ for the noncommunicating scenario, and $P_{AB} > 3/4$ and $P_{AC} > 1/2$ for the communicating scenario. Clearly, the region where both success probabilities outperform the classical bounds is substantially enlarged when communication is allowed. Nine pairs of experimentally measured success probabilities, $P_{AB}$ and $P_{AC}$, acquired for nine sharpness values $\eta$, are plotted in the same figure. All data points lie close to the optimal quantum curves, confirming that near-optimal quantum correlations are achieved across the whole range of sharpness values considered. 
Figure~\ref{fig:4} shows the differences between the upper and lower bounds of Bob's sharpness parameter $\eta$ for nine target values in both the communication-assisted and noncommunicating scenarios. Clearly, our communication strategy substantially narrows the estimation interval for $\eta$. Moreover, precision improves (i.e., the interval contracts) as $\eta$ increases.

The certification of quantum advantages requires the use of incompatible measurements. Accordingly, any violation of the classical bound by the average success probabilities $P_{AB}$ and $P_{AC}$ serves as a witness to the incompatibility of Bob's and Charlie's respective pairs of measurements \cite{tavakoli2020measurement, carmeli2020quantum}. We further quantify the corresponding measurement incompatibility, $D_B$ and $D_C$, using the nine pairs of experimentally obtained probabilities $P_{AB}$ and $P_{AC}$. The results are shown in Fig.~\ref{fig:5}. As predicted, $D_B$ decreases with decreasing $\eta$, vanishing near the theoretical threshold of $\eta = 1/\sqrt{2}$ where quantum advantage disappears. For Charlie, however, $D_C$ gradually fails to accurately capture the true degree of incompatibility as Bob's measurement sharpness increases. This is due to the sequential design of the protocol, where Bob's measurements destroy the correlation shared between Alice and Charlie.
 
\section{V. Conclusions}
In conclusion, we have proposed a DI protocol for characterizing unsharp instruments in a sequential QRAC scenario, assisted by entanglement and restricted communication. Allowing the first decoder to transmit its measurement setting to the second enables a decoding regime in which both decoders' success probabilities simultaneously exceed their classical bounds over a larger sharpness region than noncommunicating protocols. Based on the trade-off between both decoders' success probabilities, we obtain tighter bounds on the sharpness of the first decoder's instrument while simultaneously quantifying the incompatibility of both decoders' measurements. 

Experimentally, we implement the protocol on a photonic platform using a tunable Mach-Zehnder interferometer to realize the unsharp instrument. The results confirm the theoretical predictions across nine target sharpness values, approximately uniformly distributed over $[0,1]$. Compared with previous noncommunicating SDI and DI schemes in Refs.~\cite{xiao2021widening,zhou2025experimental,roy2023device}, our communication-assisted scheme consistently yields narrower sharpness-estimation intervals, with precision improving as sharpness increases. In addition, we quantify the incompatibility of the first decoder's two unsharp measurements and the second decoder's two projective measurements across the same set of target values, demonstrating both the predicted quantum advantage and the robustness of the communication-assisted protocol.

This work introduces a device-independent approach that simultaneously characterizes measurement sharpness and incompatibility. The restricted communication scenario considered here captures practical situations where parties possess different levels of contextual information---a common feature in quantum networks, including repeater protocols and distributed quantum computing, where communication resources are often limited and asymmetric. The trade-off between the two decoders' success probabilities quantitatively characterizes how such informational asymmetry can be exploited to enhance overall performance. Our protocol opens several promising directions. Extending the communication-assisted scheme to higher-dimensional systems, multipartite scenarios with more than two sequential unsharp measurements, or networked settings may enable richer forms of DI certification. Moreover, integrating adaptive measurement strategies could further tighten sharpness and incompatibility bounds under realistic noise, while also revealing new advantages in correlation sharing and randomness expansion.
 
\section{Acknowledgments}
This work was supported by the National Key Research and Development Program of China (Grant No. 2025YFE0217700), the Shandong Provincial Natural Science Foundation (Grants No. ZR2024LLZ003 and No. ZR2026LLZ016), the Fundamental Research Funds for the Central Universities (Grant  No. 202364008), and the Young Talents Project at Ocean University of China (Grant No. 861901013107).

\section*{DATA AVAILABLE}
The data that support the findings of this article are not publicly available. The data are available from the authors upon reasonable request.

\section{Appendix A: Quantifying the degree of incompatibility of Bob's pair of measurements }\label{DBApp} 
Here we present the detailed derivation of the degree of incompatibility of Bob’s pair of measurements in the scenario of an entanglement-assisted sequential $2\rightarrow 1$ QRAC with restricted communication. Alice encodes her classical input by performing a local measurement $A_{a\vert\vec{r}_{x}} = (I+(-1)^a \vec{r}_{x} \cdot \vec{\sigma})/2$ along the direction $\vec{r}_{x}$ on her half of the shared state $\rho_{AB}$, obtaining outcome $a \in \{0,1\}$. After Alice's measurement, Bob's state collapses to
\begin{equation}
\rho^B_{a\vert\vec{r}_{x}}=\dfrac{\operatorname{Tr}_A[(A_{a\vert\vec{r}_{x}}\otimes I)\rho_{AB} ]}{\operatorname{Tr}[(A_{a\vert\vec{r}_{x}}\otimes I) \rho_{AB}]}.
\end{equation}
We express Bob's conditional state in Bloch form as
\begin{equation}\label{BCS}  
\rho^B_{a\vert\vec{r}_{x}} = \frac{1}{2}(\mathbb{I} + \vec{n}^B_{a\vert\vec{r}_{x}} \cdot \vec{\sigma}),
\end{equation} 
where $\vec{n}^B_{a\vert\vec{r}_{x}}$ is the Bloch vector. Bob then performs an unsharp measurement $B_{b\vert\vec{s}_{y}} = (I+(-1)^b \vec{s}_{y} \cdot \vec{\sigma})/2$ to decode the desired input bit $x_y$. The average success probability for Bob can be written as

\begin{equation} 
P_{AB} = \dfrac{1}{2} + \dfrac{1}{8}\left[ \vec{n}^B_{0} \cdot (\vec{s}_{0}+\vec{s}_{1}) + \vec{n}^B_{1} \cdot (\vec{s}_{0}-\vec{s}_{1}) \right],
\end{equation}
where $\vec{n}^B_{x} = (\vec{n}^B_{0\vert\vec{r}_{x}} - \vec{n}^B_{1\vert\vec{r}_{x}})/2$. Since Bob's conditional state is pure, the norm of $\vec{n}^B_{x} $ s unity, i.e., $ \vert \vec{n}^B_{x}\vert =1 $.

For a given pair of measurement directions $\{\vec{s}_0,\vec{s}_1\}$, the success probability $P_{AB}$ can be maximized over Alice’s measurement choices. Clearly, Alice's optimal strategy is to align $\vec{n}^B_{0}$ with $\vec{s}_{0}+\vec{s}_{1}$ and $\vec{n}^B_{1}$ with $\vec{s}_{0}-\vec{s}_{1}$. This yields

\begin{equation}\label{maxPAB}
\begin{split}
P_{AB}^{max}=\max \limits_{\vec{r}_{x}} P_{AB} &= \dfrac{1}{2} + \dfrac{1}{8}\left( |\vec{s}_0 + \vec{s}_1| + |\vec{s}_0 - \vec{s}_1| \right), \\
&= \dfrac{1}{2}\left( 1 + \dfrac{D_B+ 2}{4} \right),
\end{split}
\end{equation}
where $D_B= |\vec{s}_0 + \vec{s}_1| + |\vec{s}_0 - \vec{s}_1| -2$ is the degree of incompatibility of Bob's pair of measurements. Rearranging Eq. (\ref{maxPAB}) gives

\begin{equation}\label{Bobincomp}
D_B \geq \text{max}{\{8 P_{AB} - 6,0}\}.
\end{equation}
Hence, by taking the observed success probability  $P_{AB}$ into Eq. \eqref{Bobincomp}, we can obtain a lower bound on the degree of incompatibility of Bob's pair of measurements. This provides an operational witness of measurement incompatibility in the entanglement-assisted sequential QRAC scenario with restricted communication.

\section{Appendix B: Quantifying the degree of incompatibility of Charlie's pair of measurements} 

Charlie's measurement incompatibility is quantified similarly to Bob's, except that Charlie's state is averaged over Bob's outcomes because Bob can communicate his measurement setting to Charlie.

Bob's unsharp instrument $B_{b\vert\vec{s}_{y}}=(I+(-1)^b \eta\vec{s}_{y}\cdot\vec{\sigma})/2$ can be expressed by Kraus operators as
\begin{equation}\label{BKS}  
\begin{aligned}
K_{0|\vec{s}_y} &= \sqrt{\tfrac{1+\eta}{2}}\,B_{0|\vec{s}_y}+\sqrt{\tfrac{1-\eta}{2}}\,B_{1|\vec{s}_y},\\
K_{1|\vec{s}_y} &= \sqrt{\tfrac{1-\eta}{2}}\,B_{0|\vec{s}_y}+\sqrt{\tfrac{1+\eta}{2}}\,B_{1|\vec{s}_y}.
\end{aligned}
\end{equation}

After Alice performs measurement $ A_{a\vert\vec{r}_{x}}$ and Bob implements an unsharp measurement along $\vec{s}_{y}$,  Charlie's conditional state becomes
\begin{equation} \label{CCS} 
\begin{split}
\rho^{C\vert\vec{s}_y}_{a\vert\vec{r}{x}} =\dfrac{1}{2}\sum \limits_{b} K_{b\vert\vec{s}_{y}}\rho^B_{a\vert\vec{r}_{x}} K^{\dagger}_{b\vert\vec{s}_{y}}.
\end{split}
\end{equation}

Substituting Eq. (\ref{BCS}) and Eq. (\ref{BKS}) into Eq. (\ref{CCS}) yields
\begin{equation}
\rho_{a|\vec{r}_x}^{C|\vec{s}_y}=\frac12\Bigl(I+\bigl[\sqrt{1-\eta^2}\,\vec{n}_{a|\vec{r}_x}^B+(1-\sqrt{1-\eta^2})(\vec{n}_{a|\vec{r}_x}^B\cdot\vec{s}_y)\vec{s}_y\bigr]\cdot\vec{\sigma}\Bigr).
\end{equation}
Charlie's conditional state can therefore be written as
\begin{equation}  
\rho^{C\vert\vec{s}_y}_{a\vert\vec{r}_{x}} = \frac{1}{2}(\mathbb{I} + \vec{n}^{C\vert\vec{s}_y}_{a\vert\vec{r}_{x}} \cdot \vec{\sigma}),
\end{equation}
with the Bloch vector
\begin{equation}  
\vec{n}_{a|\vec{r}_x}^{C|\vec{s}_y}= \sqrt{1-\eta^2}\,\vec{n}_{a|\vec{r}_x}^B+(1-\sqrt{1-\eta^2})(\vec{n}_{a|\vec{r}_x}^B\cdot\vec{s}_y)\vec{s}_y.
\end{equation}  
As discussed in Appendix A, to maximize the value of $P_{AB}$, $\vec{n}^B_{0}$ should be aligned with $\vec{s}_{0}+\vec{s}_{1}$ and $\vec{n}^B_{1}$ with $\vec{s}_{0}-\vec{s}_{1}$. Under this optimal choice, one finds $|\vec{n}_{a|\vec{r}_x}^{C|\vec{s}_y}|=\sqrt{1-\eta^2/2}$.


Charlie then performs a projective measurement $C_{c|\vec{t}_z}=[I+(-1)^c\vec{t}_z\cdot\vec{\sigma}]/2$  to decode the desired input bit $x_z$. The average success probability for Charlie can be written as

\begin{equation} 
P_{AC} = \dfrac{1}{2} + \dfrac{1}{8}\left[ \vec{n}^C_{0} \cdot (\vec{t}_{0}+\vec{t}_{1}) + \vec{n}^C_{1} \cdot (\vec{t}_{0}-\vec{t}_{1}) \right],
\end{equation}
where $\vec{n}^C_{x} = (\vec{n}_{0|\vec{r}_x}^{C|\vec{s}_y} - \vec{n}_{1|\vec{r}_x}^{C|\vec{s}_y})/2$. Following the optimization procedure outlined in Appendix A, the maximum success probability for Charlie is
\begin{equation}\label{maxPAC} 
\begin{split}
P_{AC}^{\max}&= \frac12+\dfrac{|\vec{n}_{a|\vec{r}_x}^{C|\vec{s}_y}|}{8}\bigl(|\vec{t}_0+\vec{t}_1|+|\vec{t}_0-\vec{t}_1|\bigr),\\
&= \frac12+\dfrac{|\vec{n}_{a|\vec{r}_x}^{C|\vec{s}_y}|}{8}\bigl(D_C +2\bigr), 
\end{split}
\end{equation}  
where $ D_C = |\vec{t}_0+\vec{t}_1|+|\vec{t}_0-\vec{t}_1|-2 $ is the degree of incompatibility of Charlie's pair of measurements.  The maximum is achieved when $\vec{n}^C_{0}$ is aligned with $\vec{t}_{0}+\vec{t}_{1}$ and $\vec{n}^C_{1}$ with $\vec{t}_{0}-\vec{t}_{1}$. The norms of both $\vec{n}^C_{0}$  and $\vec{n}^C_{1}$  are bound by $|\vec{n}_{a|\vec{r}_x}^{C|\vec{s}_y}|$.  Rearranging Eq. (\ref{maxPAC}) yields
\begin{equation}\label{BoundDC} 
\begin{split}
D_C\ge  \dfrac{8P_{AC}-4}{\sqrt{1-\eta^2/2}}-2. 
\end{split}
\end{equation} 
Clearly, the right-hand side of Eq. (\ref{BoundDC}) is an increasing function of $\eta$. As mentioned in the main text
\begin{equation}\label{Boundeta}
	\begin{aligned}
	 \sqrt{2}(2P_{AB}-1)\leq \eta \leq \sqrt{8P_{AC}-8(P_{AC})^2-1}.
		\end{aligned}
\end{equation}
The tightest (largest) lower bound of $D_C$ is obtained by taking the maximum possible value of  $\eta$, i.e.,
\begin{equation}  
D_C \ge \rm{max}{\{ \dfrac{8P_{AC}-4}{\sqrt{4P_{AC}^2-4P_{AC}+3/2}}-2,0 }\}.
\end{equation}  
This lower bound is clearly independent of $P_{AB}$, in contrast to the case without communication {\cite{anwer2020experimental}.

\end{document}